# Title: Quasicrystalline oxide interfaces by twisting $SrTiO_3$ membranes

**Authors:** Shivasheesh Varshney[1,*], Sooho Choo[1], Huan Liu[2], In Hyeok Choi[3], Nicolas Gauquelin[6], John Mangeri[5], Seung Gyo Jeong[1,4], Edwin Dollekamp[5], William Sandholt[5], Andrea Roberto Insinga[5], Nikolas Vitaliti[5], Bridget R. Denzer[7], Bonnie Y. X. Lin[7], Andrew T. Danbury[7], Majid Mohseni[1], Uditha Mihiranga Jayathilake[8], Sharup Sheikh[8], Lidia A. Lapinski[8], Jonas M. Hansen[5], Seungjun Lee[11,12], Dong Kyu Lee[1], Jitin Sathish Kumar[1], Shivam Sharma[9], Rishi Raj[1], Jay Shah[1], Alevtina Smekhova[10], Florian Kronast[10], Juan Maria García Lastra[5], K. Andre Mkhoyan[1], Tony Low[11], James M. LeBeau[7], Alexander X. Gray[8], Jo Verbeek[6], Keith A. Nelson[3], Nini Pryds[5], Richard D. James[13], and Bharat Jalan[1,*]

**Affiliations:**

[1] Department of Chemical Engineering and Materials Science, University of Minnesota, Minneapolis, Minnesota, 55455, USA
[2] Mechanical & Civil Engineering, California Institute of Technology, Pasadena, California, 91125, USA
[3] Department of Chemistry, Massachusetts Institute of Technology, Cambridge, Massachusetts 02139, USA
[4] Department of Physics, Hankuk University of Foreign Studies, Yongin, 17035, Republic of Korea
[5] Department of Energy Conversion and Storage, Technical University of Denmark, Anker Engelunds Vej 1, Kongens Lyngby 2800, Denmark
[6] EMAT and Nanolight Center of Excellence, Department of Physics, University of Antwerpen, Groenenborgerlaan 171, 2020 Antwerpen, Belgium
[7] Department of Materials Science and Engineering, Massachusetts Institute of Technology, Cambridge, Massachusetts, 02139, USA
[8] Department of Physics, Temple University, Philadelphia, Pennsylvania, 19122, USA
[9] Material Research Laboratory, University of Illinois, Urbana-Champaign, Illinois, 61801, USA
[10] Helmholtz-Zentrum Berlin für Materialien und Energie GmbH, Berlin, 12489, Germany
[11] Department of Electrical and Computer Engineering, University of Minnesota, Minneapolis, Minnesota 55455, USA
[12] Department of Applied Physics, Kyung Hee University, Yongin, 17104, Republic of Korea
[13] Department of Aerospace Engineering and Mechanics, University of Minnesota, Minneapolis, Minnesota, 55455, USA

*Corresponding authors: varsh022@umn.edu ; bjalan@umn.edu

**Abstract:**

Quasicrystals reveal that long-range order can exist without translational periodicity, but deterministic routes for designing strongly bonded quasicrystalline interfaces remain unexplored. Here, we show that such interfaces can be engineered by growing epitaxial $SrTiO_3$ membranes on a sacrificial layer, releasing them, twisting them by prescribed angles, and bonding them onto a chosen substrate. $SrTiO_3$ bilayers twisted by 45° form an octagonal quasicrystalline interface with eight-fold rotational symmetry, whereas trilayers assembled with successive 30° rotations form a dodecagonal interface with twelve-fold symmetry. Selected-area and nano-beam electron diffraction, multislice electron ptychography, atomic-resolution dark field imaging, Fourier analysis, and real-space tiling establish long-range quasicrystalline order with electronic reconstruction evident from electron energy-loss spectroscopy after annealing. Atomistic simulations based on machine-learned interatomic potentials show that octagonal order emerges as the bilayer twist angle approaches 45°. Polarization-dependent X-ray absorption spectroscopy in 1.5 nm twisted bilayers reveals a twist-dependent electronic response, and terahertz-field-induced second-harmonic generation circular dichroism identifies a symmetry-dependent nonlinear optical response at the 45° quasicrystalline condition. Twisted oxide membranes therefore provide an extendable route to engineer quasiperiodic order and couple it to interfacial bonding, orbital reconstruction, and nonlinear optical response in complex oxides.

**Main Text**

Quasicrystals established that long-range order can exist without translational periodicity (*1*, *2*). Their atomic arrangements are deterministic rather than random, giving rise to sharp diffraction patterns with rotational symmetries–including 5-fold, 8-fold, 10-fold, and 12-fold symmetry–that cannot occur in periodic crystals (*3*, *4*). This form of order has been linked to unusual electronic, optical, mechanical, and topological responses (*5*–*9*). Yet most quasicrystals have been discovered in specific alloy chemistries or stabilized as ultrathin phases on particular substrates (Fig. S1), thereby limiting deliberate control over their symmetry, interface chemistry, dimensionality, and integration.

A route to designing quasicrystalline matter is to make the rotation angle itself the control parameter. Recent advances in epitaxial growth and freestanding membrane release now make this strategy experimentally accessible in crystalline thin films (*10*–*17*). Twisted van der Waals systems have shown that forbidden rotational symmetries can be imposed geometrically (*6*, *18*, *19*). In those materials, however, the layers remain weakly coupled, and quasiperiodicity is introduced mainly through projected registry between otherwise intact two-dimensional (2D) crystals (*18*, *20*–*23*). Complex oxides offer a distinct opportunity: their exposed ionic-covalent surfaces can bond, reconstruct, and modify the local electronic structure upon contact (*24*). A twisted oxide interface can therefore be more than a projected quasicrystalline pattern; it can become a chemically bonded and electronically reconstructed quasicrystalline boundary.

Here we demonstrate this concept using single-crystalline $SrTiO_3$ (STO) membranes. Epitaxial STO membranes are released from sacrificial layers, rotated by prescribed angles, stacked, and annealed to form bonded twist interfaces on chosen substrates. A 45° twisted bilayer forms an octagonal quasicrystalline interface with 8-fold rotational symmetry, whereas a trilayer with a sufficiently thin middle layer assembled with successive 30° rotations forms a dodecagonal quasicrystalline interface with 12-fold symmetry. Electron diffraction, multislice electron ptychography (MEP), high angle annular dark field (HAADF) imaging, reciprocal-space analysis, and real-space tiling establish long-range quasiperiodic order. Atomistic simulations using machine-learned interatomic potentials (MLIP) further show that octagonal order emerges as the bilayer twist angle approaches 45°, distinguishing the quasicrystalline state from nearby commensurate approximants. By separating quasicrystalline order from the growth substrate, this membrane-based approach avoids any potential substrate-related contributions, including those from metallic supports that can dominate electronic structure and limit integration in substrate-attached oxide quasicrystals (*10–17*). In contrast, freestanding STO membranes allow the quasicrystalline boundary to form after growth and to be transferred onto different supports. Because lattice, charge, orbital, and spin degrees of freedom are strongly coupled in complex oxides, such bonded twist interfaces may provide a route to engineer not only forbidden rotational symmetry, but also local electronic and optical responses at a designed boundary.

We test this idea using complementary structural, spectroscopic, and nonlinear optical probes. Electron energy loss spectroscopy (EELS) shows that annealing produces a strongly bonded twist interface accompanied by local charge redistribution. Polarization-dependent X-ray absorption spectroscopy (XAS) at the O *K*-edge and Ti $L_{2,3}$-edge reveals site-selective orbital reconstruction in 1.5 nm 45° twisted bilayers. Terahertz-field-induced second-harmonic generation circular dichroism (SHG-CD) identifies the 45° twist interface as a distinct high-symmetry point in the rotational phase diagram. These results establish twisted oxide membranes as a platform for designing quasicrystalline interfaces and coupling quasiperiodic order to interfacial bonding, orbital reconstruction, and nonlinear optical response in complex oxides.

**Results and discussion**

The assembly of freestanding, single-crystalline oxide membranes provides a direct route to quasicrystalline interfaces. In this approach, each STO membrane acts as a square-lattice building block with fourfold rotational symmetry (Fig. 1A). Because cubic STO has an in-plane lattice constant of 3.905 Å, one membrane can be twisted relative to another to create a well-defined superposition of square lattices without requiring a specific growth substrate (Figs. S2 and S3). A bilayer twisted by 45° produces an octagonal pattern with 8-fold rotational symmetry. In contrast, a trilayer assembled with successive 30° rotations produce a dodecagonal pattern with 12-fold rotational symmetry (Fig. 1B). These two twist configurations realize the canonical 2D octagonal and dodecagonal quasicrystalline symmetries in an oxide membrane platform. Because the Fourier transform is linear, the superposition of the rotated lattices also superposes their diffraction patterns, producing discrete rotational symmetries (8-fold and 12-fold) that are not possible in periodic crystals.

Precise angular control is central to this strategy. X-ray diffraction $\phi$-scans of the STO (103) reflection on $Al_2O_3$, which resolve the in-plane crystallographic orientation of each membrane, show that the fabricated twist angles match the target values within ± 0.2° over the full range from 0° to 45° (Figs. 1C, D). The same angular precision is retained on $Al_2O_3$, $SiO_2$/Si, Si, and STO substrates, demonstrating that the twist is defined by membrane assembly rather than by the support (Fig. S4). Diffraction peaks separated by the prescribed angles, including 12.8°, 25.2°, 36.2°, and 45.2°, confirm the superposition of independently oriented single-crystalline STO layers. Thus, the transition from the native 4-fold symmetry of an individual STO membrane to 8-fold symmetry in the twisted bilayer is imposed geometrically by the twist angle. The released membranes expose SrO-terminated interfacial surfaces formed during growth, producing a chemically distinct twist boundary (Fig. S4). Laue fringes in $2\theta$-$\omega$ scans, Kiessig fringes in grazing-incidence X-ray reflectivity, and correlated SEM/AFM measurements confirm smooth, continuous bilayers with individual layer thicknesses of 31 and 32 nm and sub-nanometer surface roughness after stacking (Figs. S4 and S5).

We next verified quasicrystalline order in reciprocal and real space. MEP on 10 nm/10 nm 45° twisted bilayers resolves the square STO lattice from both the bottom and top layers as well as their overlap by averaging a 4 nm-region through the depth at the boundary between top and bottom layers (Fig. 2A). Nano-beam electron diffraction (NBED) from the 45° bilayer shows eight equally spaced diffraction spots on each reciprocal-space shell. In contrast, selected-area electron diffraction (SAED) from the 30°/30° trilayer shows twelve equally spaced spots (Fig. 2B). The diffraction peaks are sharp, indicating long-range orientational order rather than rotational disorder. In the 12-fold trilayer, weaker spots are from the 3.12 nm thick bottom layer. All layers are visible, highlighting the layer-resolved nature of the quasicrystalline projection. Plane-view HAADF-STEM resolves the corresponding quasiperiodic atomic arrangements in real space (Fig. 2C). Fourier transforms of these images reproduce the 8-fold and 12-fold diffraction symmetries through multiple octagonal and dodecagonal shells (Fig. 2D), linking the forbidden rotational symmetry in reciprocal space directly to the real-space quasiperiodic lattice. Additional satellite features appear in the Fourier transform relative to the idealized pattern in Fig. 1B, likely reflecting the finite thickness and projected depth structure of the membrane stack.

Using MLIPs, we tested whether the quasicrystalline interface is preserved after atomic relaxation. Because exact 45° quasiperiodicity cannot be represented in a finite periodic supercell, simulations were performed on a commensurate approximant with a twist angle of 44.76°. MLIP-relaxed structures at 0 K and molecular-dynamics (MD) configurations at 300 K were examined for both SrO-$TiO_2$ and SrO-SrO terminations (Figs. S6 and S7). In all cases, the interfacial arrangement remains intact, with only modest local displacements. Thus, the local motif near the 45° condition is not erased by relaxation or thermal motion. This behavior contrasts with low twist angles in STO, where atomic relaxation can drive pronounced reconstruction toward lower-energy commensurate configurations (*24*).

To verify that the atomic contrast in Fig. 2C reflects the underlying quasicrystalline lattice, we performed *ab initio* multislice STEM image simulations (abTEM) using MLIP-relaxed atomic structures for the 44.76° twisted bilayer. Simulations of rigid SrO-$TiO_2$-terminated structures (Fig. S8) and relaxed SrO-$TiO_2$ and SrO-SrO terminations (Figs. S9 and S10) produce nearly identical quasiperiodic contrast patterns, showing that the observed motif is set primarily by rotational geometry rather than by a particular termination. The simulated images reproduce the experimental plane-view ptychography contrast of the bilayer (Fig. S11), confirming that the pattern arises from the projected atomic arrangement of the twisted STO lattice. For the trilayer, a fully MLIP-relaxed three-layer simulation was computationally prohibitive. We therefore simulated a rigid 30°/30° trilayer interface, which closely matches the experimental trilayer image and supports the structural assignment of the observed 12-fold pattern (Figs. S12 and S13). Fast Fourier transforms of the plane-view STEM images reproduce the octagonal and dodecagonal symmetries, confirming that the real-space atomic arrangements generate the 8-fold and 12-fold rotational symmetries (Fig. 2D). The origin of these symmetries is further captured by the cut-and-project construction, which is provided in the Supplementary Materials (Fig. S14, S15 and Note S7) (*25*, *26*).

To distinguish quasicrystalline order from nearby commensurate approximants, we calculated the two-dimensional structure factor, $S(\mathbf{q}) = N^{-1}|\Sigma_j \exp(i\mathbf{q}\cdot\mathbf{r}_j)|^2$, from simulated atomic configurations. We then evaluated a composite quasicrystalline order parameter, Q, that captures 8-fold angular order and the agreement of radial shell positions with the ideal octagonal sequence (*27*). Integer shell ratios, which can occur in periodic crystals, were assigned no diagnostic weight, whereas irrational ratios characteristic of octagonal order were weighted more strongly (Fig. S16). The order parameter approaches Q = 1.0 at $\Delta\phi = 45.0°$, the ideal quasicrystalline condition, while the closest available relaxed commensurate approximant at $\Delta\phi = 44.76°$ gives Q = 0.92.

We then used real-space tiling analysis to connect the observed atomic arrangements to canonical mathematical quasicrystals. Large-area plane-view HAADF-STEM images of the bilayer and trilayer are shown in Figs. S17 and S18, with ideal tiling overlays in Fig. S19. For the 8-fold bilayer, the atomic motifs map onto the Ammann-Beenker tiling, a two-tile octagonal quasicrystal composed of squares and 45° rhombi with a common edge length (Fig. S20A) (*28*). The sides of individual tiles are oriented in exactly eight distinct directions separated by 45°. The square-to-rhombus frequency ratio agrees with the theoretical value $1:\sqrt{2}$. For the 12-fold trilayer, the atomic positions map onto the Stampfli tiling, a dodecagonal quasicrystal composed of equilateral triangles, squares, and 30° rhombi with a common edge length (Fig. S20B) (*29*). The tile edges occur in twelve orientations separated by 30°, and tile counting gives the expected frequency ratio N(triangle):N(square):N(rhombus) = $\sqrt{3}:1:\sqrt{3}$. These real-space tiling relationships independently verify that the twisted STO bilayer and trilayer are physical realizations of octagonal and dodecagonal quasicrystals.

Long-range quasiperiodic order is further reflected in the self-similar hierarchy of real-space supertiles (Figs. 2E,F). In the Ammann-Beenker bilayer (*28*), the measured cluster diameters scale as $a$, $a(1+\sqrt{2}) \approx 2.414a$, and $a(1+\sqrt{2})2 \approx 5.828a$, corresponding to successive generations related by the silver-ratio inflation factor $\delta = 1+\sqrt{2}$ (Fig. 2E) (*30*). In the Stampfli trilayer, the corresponding cluster diameters scale as $a$, $a(2+\sqrt{3}) \approx 3.732a$, and $a(2+\sqrt{3})2 \approx 13.928a$, in agreement with the dodecagonal inflation factor $\beta = 2+\sqrt{3}$ (Fig. 2F) (*29*). These inflation hierarchies provide a real-space signature of quasiperiodicity beyond local rotational symmetry.

To quantify this self-similarity in the 45° twisted bilayer, we performed an automated tiling analysis on simulated HAADF-STEM images of the ideal structure (Figs. S21-S23). The analysis used a SrO-SrO-terminated bilayer image focused on the interface, with a beam sampling rate of 0.5 $Å^{-1}$, and centered on a high-symmetry region. Ammann-Beenker tilings with different tile edge lengths were overlaid on the image, and a mismatch cost function was minimized to identify the length scales at which equivalent tile motifs best coincide with the simulated atomic contrast. The resulting local minima occur at tile sizes related by the expected silver-ratio scaling, confirming the self-similar hierarchy of the 8-fold quasicrystalline lattice (Fig. S22). When the same algorithm is applied across twist-angle approximants, the preference for irrational octagonal self-similarity is recovered only near the quasicrystalline condition (Fig. S24). A similar analysis for the trilayer confirms the expected 12-fold scaling laws (Fig. S23). This analysis provides a quantitative complement to the direct tiling and reciprocal-space measurements.

Cross-sectional HAADF-STEM confirmed that the twisted membranes form a clean, atomically abrupt interface after assembly and annealing. In the 45° twisted bilayer on Si(001), the two STO layers have thicknesses of ~ 30 nm each, consistent with the nominal MBE growth rate (Fig. S25). Before annealing, the stacked membranes contain a 1-3 nm interfacial residue layer attributed to the transfer process (Fig. S26). Annealing at 900°C in flowing $O_2$ removes this layer and brings the two crystalline membranes into direct contact (Fig. 3A). Atomic-resolution imaging shows that the twist boundary is abrupt over approximately one unit cell, with no detectable amorphous interlayer or residual contamination. The fast Fourier transform of the cross-sectional image shows [110] reflections from the upper layer and [100] reflections from the lower layer, offset by 45°, confirming that each membrane retains its crystalline orientation across the bonded twist interface (Fig. S25). An analogous cross-section of the annealed 30°/30° twisted trilayer, composed of three 5 nm STO membranes on $SiO_2$/ Si(001), confirms that the layers remain in close crystalline contact after assembly (Fig. 3B and Fig. S27).

To test whether annealing also produces strong interfacial bonding, we measured EELS across annealed and unannealed bilayer interfaces. Fig. 3C shows EELS spectra and a STEM dark-field line profile across a 45° twisted membrane transferred onto an STO (001) substrate. The line profile from the annealed SrO-SrO-terminated bilayer shows continuous lattice fringes across the twist interface with a spacing of 3.905 Å. In contrast, the same sample before annealing lacks continuous fringes across the interfacial gap (Fig. S28). The Ti $L_{2,3}$-edge EELS spectra in the top and bottom layers show the characteristic four-peak crystal-field-split $Ti^{4+}$ fine structure ($L_3$ $t_{2g}/e_g$ at ~458.7/461.0 eV and $L_2$ $t_{2g}/e_g$ at ~464.1/466.5 eV), matching the bulk $Ti^{4+}$ reference. At the twist interface, the spectra show an additional $Ti^{3+}$ contribution together with $Ti^{4+}$, consistent with local charge redistribution associated with bonding after annealing. The unannealed interface does not show an obvious $Ti^{3+}$ signal (Fig. S29) (*31*, *32*). These results underscore the role of annealing in forming a bonded oxide twist boundary.

We next investigated whether the quasicrystalline twist boundary also produces electronic reconstruction. XAS and X-ray linear dichroism (XLD), measured by X-ray Photoemission Electron Microscopy (XPEEM), were performed on samples containing spatially adjacent monolayer STO and 45° twisted bilayer STO regions (Fig. 4A and Fig. S30). This geometry enables a simultaneous internal comparison between a single 1.5 nm STO membrane and a 1.5 nm/1.5 nm 45° twisted bilayer within the same XPEEM field of view under identical measurement conditions. Thus, the primary intentional difference between the two regions is the presence of the twist interface. Spectra were acquired at a grazing incidence angle of 16° relative to the sample surface using polarization geometries sensitive predominantly to in-plane orbital character (E∥*ab*) and out-of-plane orbital character (E∥*c*). This field-of-view-based simultaneous acquisition minimizes artifacts associated with beam exposure, sample history, and substrate effects. Because the samples were annealed in $O_2$ before measurement, the dichroic changes are attributed primarily to the twist interface rather than to uncontrolled oxygen deficiency.

The Ti $L_{2,3}$-edge XLD directly probes the anisotropy of unoccupied Ti 3*d* states (Fig. S30). Relative to the monolayer reference, the twisted bilayer shows a redistribution of Ti 3*d* spectral weight toward enhanced in-plane orbital character, consistent with a modified local crystal-field environment at the twist interface. At the O *K*-edge, the O 2*p*-Ti 3*d* hybridization feature near 531 eV remains largely preserved, with a modest enhancement of out-of-plane spectral weight in the bilayer. By contrast, the O 2*p*-Sr 4*d* hybridization feature near 535 eV increases substantially in the bilayer, consistent with enhanced interfacial bonding across the twist boundary (Fig. 4A). The Ti $L_{2,3}$ and O *K*-edge spectra indicate that the 45° quasicrystalline interface is not a passive structural overlap of two STO lattices, but produces a site-selective orbital reconstruction and bonding configuration absent in isolated STO membranes.

To understand the origin of the spectroscopic changes, we performed first-principles calculations on a commensurate bilayer at $\Delta\phi = 43.6°$, close to the ideal 45° quasicrystalline condition (Fig. S31A-B). The calculated electronic structure shows dispersive conduction bands derived primarily from Ti 3*d* states and relatively flat valence bands derived from O 2*p* states. Structural relaxation also produces a redistribution of Ti-O dipoles at the interface, forming a vortex-antivortex pattern (Fig. S31C). This interfacial polarization texture is consistent with the reduced local symmetry of the twisted boundary and provides a plausible microscopic origin for the orbital reconstruction observed by XAS. Spectroscopic ellipsometry on 2 nm/2 nm STO bilayers twisted by 45° on sapphire shows an increase in the refractive index after annealing, consistent with the removal of the interfacial gap layer and the formation of a denser, bonded interface (Fig. S32). The XAS/XLD, DFT, and ellipsometry results show that the 45° quasicrystal boundary is not simply a geometric overlap of two crystals; it is a chemically bonded interface that reconstructs the local orbital and charge environment. The corresponding electronic reconstruction in 30°/30° trilayers remains an important direction for future work.

Having established the structural identity of the oxide quasicrystal interfaces, we probed their macroscopic electromagnetic response using SHG-CD, a symmetry-sensitive nonlinear optical technique (Fig. 4B). In this measurement, a high-field THz pulse drives an interfacial nonlinear response, while circularly polarized 800 nm probe pulses measure the induced second-harmonic signal. The difference between the responses to right- and left-circularly polarized probe light yields a circular-dichroic SHG signal sensitive to the symmetry of the buried interface. A detailed description of the measurement geometry and nonlinear response function is provided in the Supplementary Materials.

Fig. 4C shows the twist-angle-dependent SHG-CD response measured at 45° oblique incidence from the bare Nb:STO (001) substrate and from twisted STO membranes on Nb:STO (001). Introducing an STO membrane creates an additional buried STO/STO interface, so the twisted heterostructures exhibit TFISH-CD  responses that differ from the bare substrate, even at $\Delta\phi = 0°$. At intermediate twist angles, including 22.5°, the phase shifts relative to the substrate, indicating an interfacial contribution from modified boundary conditions (Fig. 4D). At the 8-fold quasicrystalline condition, $\Delta\phi = 45°$, the TFISH-CD phase returns to zero and the signal becomes indistinguishable from that of the substrate, indicating symmetry-dependent suppression of the interfacial nonlinear response. This response appears only after annealing (Fig. S33), highlighting the role of coherent bonding in forming the functional twist boundary. The cancellation at 45° identifies the quasicrystalline twist as a special high-symmetry point in the rotational phase diagram and links the forbidden rotational order to a macroscopic nonlinear optical response.

**Conclusion**

The twisted oxide membrane platform opens a route to quasicrystal order in a wide variety of materials where lattice, charge, orbital, and spin degrees of freedom are strongly coupled (*19*, *24*, *32–36*). Unlike substrate-attached oxide quasicrystals, where metallic supports can dominate the electronic structure, freestanding membranes isolate the twist interface and allow quasiperiodicity to be imposed by a controllable geometric parameter. The orbital reconstruction observed here shows that quasicrystal rotational symmetry can modify the local electronic structure without changing the bulk chemistry of the constituent layers. More broadly, twisted oxide membranes provide a pathway to design forbidden symmetry in functional materials and to explore electronic, optical, and collective phenomena at the intersection of quasiperiodicity and strong correlations.

**Acknowledgments:**

**Funding:**

This work is primarily supported from the Air Force Office of Scientific Research Multi University Research Initiative (AFOSR MURI, Award No. FA9550-25-1-0262). Synthesis and characterization of single layer membranes were supported by the U.S. Department of Energy through grant no. DE-SC0020211. **S.V.** also acknowledges the University of Minnesota Doctoral Dissertation Fellowship. Film growth was performed using instrumentation funded by AFOSR DURIP awards FA9550-18-1-0294 and FA9550-23-1-0085. **S.C.** and **B.J.** acknowledge partial support from the Air Force Office of Scientific Research (AFOSR) through Grant No FA9550-21-0460. **W.S.** and **J.M.** acknowledge resources on the Niflheim supercomputer at the Technical University of Denmark supported by the Novo Nordisk Foundation Data Science Research Infrastructure 2022 Grant: A high-performance computing infrastructure for data-driven research on sustainable energy materials, grant no. NNF22OC0078009. **H.L.**, **S.S.**, and **R.D.J.** also benefitted from the support of AFOSR via #FA9550-24-1-0344. **S.G.J.** was partly supported by the U.S. Department of Energy through grant No. DE-SC0024710. **S.G.J** also acknowledges support from Hankuk University of Foreign Studies Research Fund. **I.H.C.** was supported by the National Research Foundation of Korea (NRF) grant funded by the Korea government (MSIT) (No. RS-2024-00351794). This work was supported in part by the Office of Naval Research (grant no. 14035670). **B.R.D.**, **B.Y.X.L**, **A.T.D**, and **J.M.L.** acknowledge support from the AFOSR through grant No. FA9550-24-1-0266. **B.R.D.** acknowledges the support of the National Science Foundation Graduate Research Fellowship under Grant No. 2141064. **A.X.G.** also gratefully acknowledges the support from the Alexander von Humboldt Foundation. **S.L.** acknowledges support of the Basic Science Research Program through the National Research Foundation of Korea funded by the Ministry of Education (NRF-2021R1A6A3A14038837). Early brainstorming on this line of research benefitted from a Vannevar Bush Faculty Fellowship. N. G. and J. V. acknowledge funding from the EU's Horizon Europe research and innovation program under grant agreement no. 101130652 (RIANA). **N.P.** acknowledges the support by the ERC Advanced (NEXUS, grant no. 101054572), by a research grant (VIL73726) from Villum Fonden and by ERC Synergy Grant METRIQS. Parts of this work were carried out in the Characterization Facility, University of Minnesota, which receives partial support from the NSF through the MRSEC (Award Number DMR-2011401) and the NNCI (Award Number ECCS-2025124) programs. Portions of this work were conducted in the Minnesota Nano Center, which is supported by the National Science Foundation through the National Nanotechnology Coordinated Infrastructure (NNCI) under Award Number ECCS-2025124. The authors thank the Helmholtz-Zentrum Berlin für Materialien und Energie for allocation of synchrotron radiation beamtime on beamline UE49-PGM-a at the BESSY II electron storage ring (232-12354-ST and 242-12681-ST).

**Author contributions:**

Conceptualization: BJ, SV, RDJ
Methodology: SV, SC, JM, ARI, AXG, NP, NG, and BJ
Investigation: SV, HL, UMJ, SS, LAL, AS, FK, WSH, SGJ, BD, BL, NG
Visualization: HL, JM, ARI, IHC, NG
Funding acquisition: BJ, NP, RDJ, KAN, TL, AXG, KAM, JMGL, JML
Project administration: BJ
Supervision: BJ
Writing – original draft: SV, HL, NG, BJ
Writing – review & editing: SV, SC, HL, IHC, ED, NV, BRD, BYXL, ATD, NG, JM, WSH, ARI, MM, UMJ, SS, SGJ, JMH, SL, DKL, JSK, SS, RR, AS, FK, JMGL, KAM, TL, JML, KAN, JV, AXG, RDJ, NP, and BJ

**B.J., S.V.** and **R.D.J.** conceived the idea and directed the project. **S.V. R.D.J.** and **B.J.** wrote the manuscript. **S.V.** grew films for bilayers using MBE, exfoliated membranes, and fabricated twisted bilayers under supervision of **B.J. N.V.** grew films for trilayer using PLD, exfoliated membranes, and **E.D.** fabricated twisted trilayers under supervision of **N.P. H.L.** and **S.S.** worked on theory for quasicrystal under the supervision of **R.D.J. J.M.** performed quasicrystalline factor simulations under the supervision of **N.P. W.S.**, **J.M.**, **J.M.H.** performed MLIP and abTEM simulations under the supervision of **J.M.G.** and **N.P. A.R.I**., **W.S.** and **J.M.** performed self-similarity analysis on real-space image tiling and atomistic diffraction spectra. **B.R.D.**, **B.Y.X.L., and A.T.D.** performed electron ptychography imaging under the supervision of **J.M.L. U.M.J.**, **S.S.,** and **L.A.L.** performed XAS/XLD under supervision of **A.X.G.**, and in collaboration with **A.S.** and **F.K. N.G.** performed cross-section STEM imaging for bilayer and trilayer. **N.G.** performed cross-section imaging, EELS measurements and HAADF plane-view imaging for bilayer and trilayers. **R.R.** performed plane-view imaging for bilayer under the supervision of **K.A.M. S.G.J.** and **S.V.** performed ellipsometry; **I.H.C.** performed Tera-hertz field induced SHG-CD under the supervision of **K.A.N. S.L.** performed electronic structure calculations under the supervision of **T.L.** All authors contributed to discussion.

**Competing interests:** Authors declare that they have no competing interests.

**Data and materials availability:** All data are available in the main text or the supplementary materials.

**List of Supplementary Materials**

Materials and Methods

Supplementary Text (Notes S1–S14)

Figs. S1 to S33

References (*39–92*)

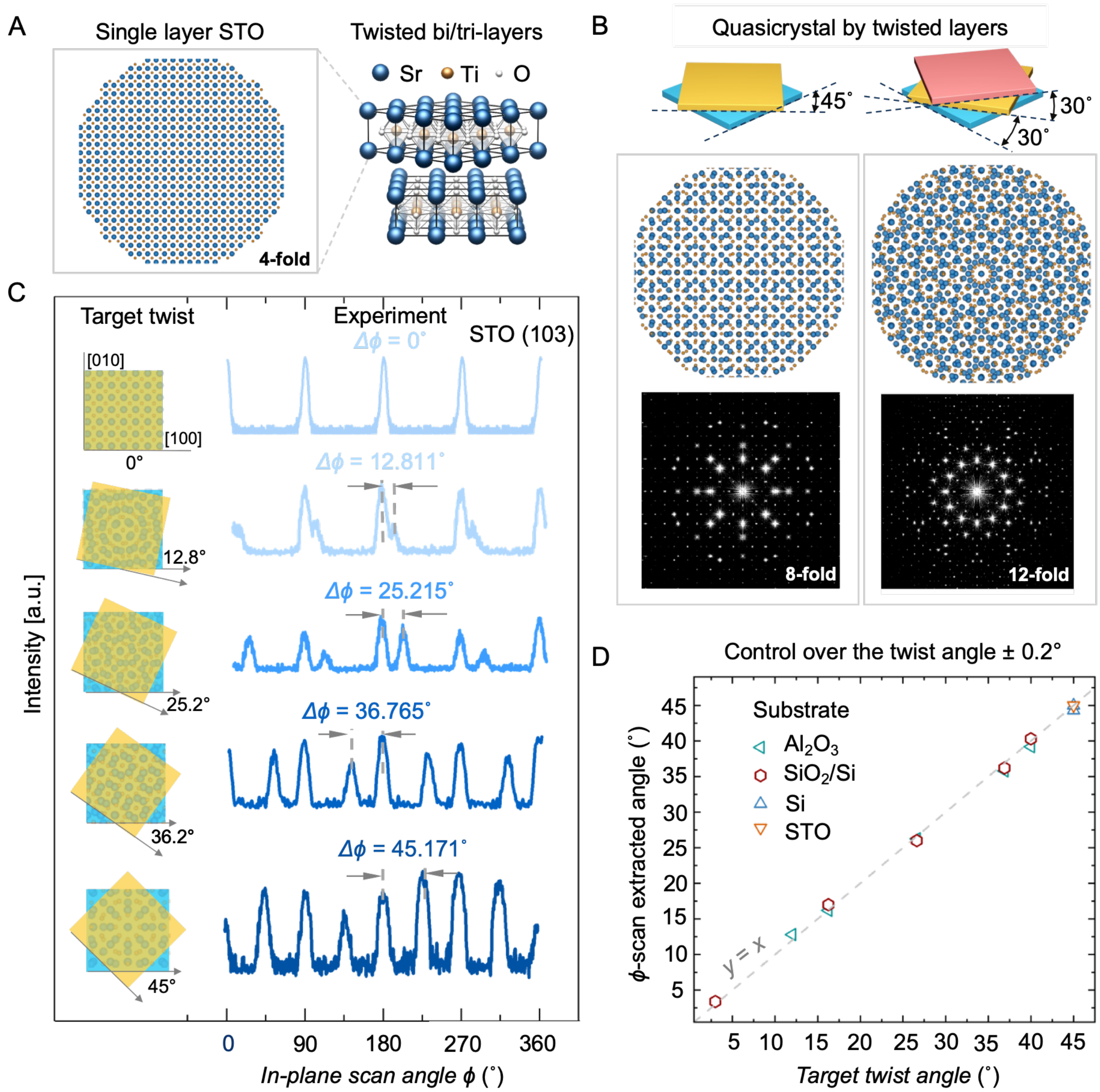

**Fig. 1.** Twist-defined formation of octagonal and dodecagonal STO quasicrystalline interfaces. (**A**) Plane-view image of a single STO membrane showing the native 4-fold symmetry. The inset shows a cross-sectional schematic of a twisted bilayer; Sr, Ti, and O atoms are shown in blue, yellow, and gray, respectively. (**B**) Schematics of a 45° twisted bilayer and a 30°/30° twisted trilayer, with corresponding plane-view images and simulated Fourier transforms showing 8-fold and 12-fold symmetries (with filtered noise). (**C**) X-ray diffraction in-plane $\phi$-scans around the STO (103) reflection for a 30 nm STO membrane stacked on a second 30 nm STO membrane on $Al_2O_3$ substrate, showing twist angles of 0°, 12.811°, 25.215°, 36.765°, and 45.171°. (**D**) Extracted twist angle from $\phi$-scans plotted against the target twist angle on various host substrates. Instrumental angular uncertainty is ± 0.002°; the sample-to-sample twist precision is ±0.2°.

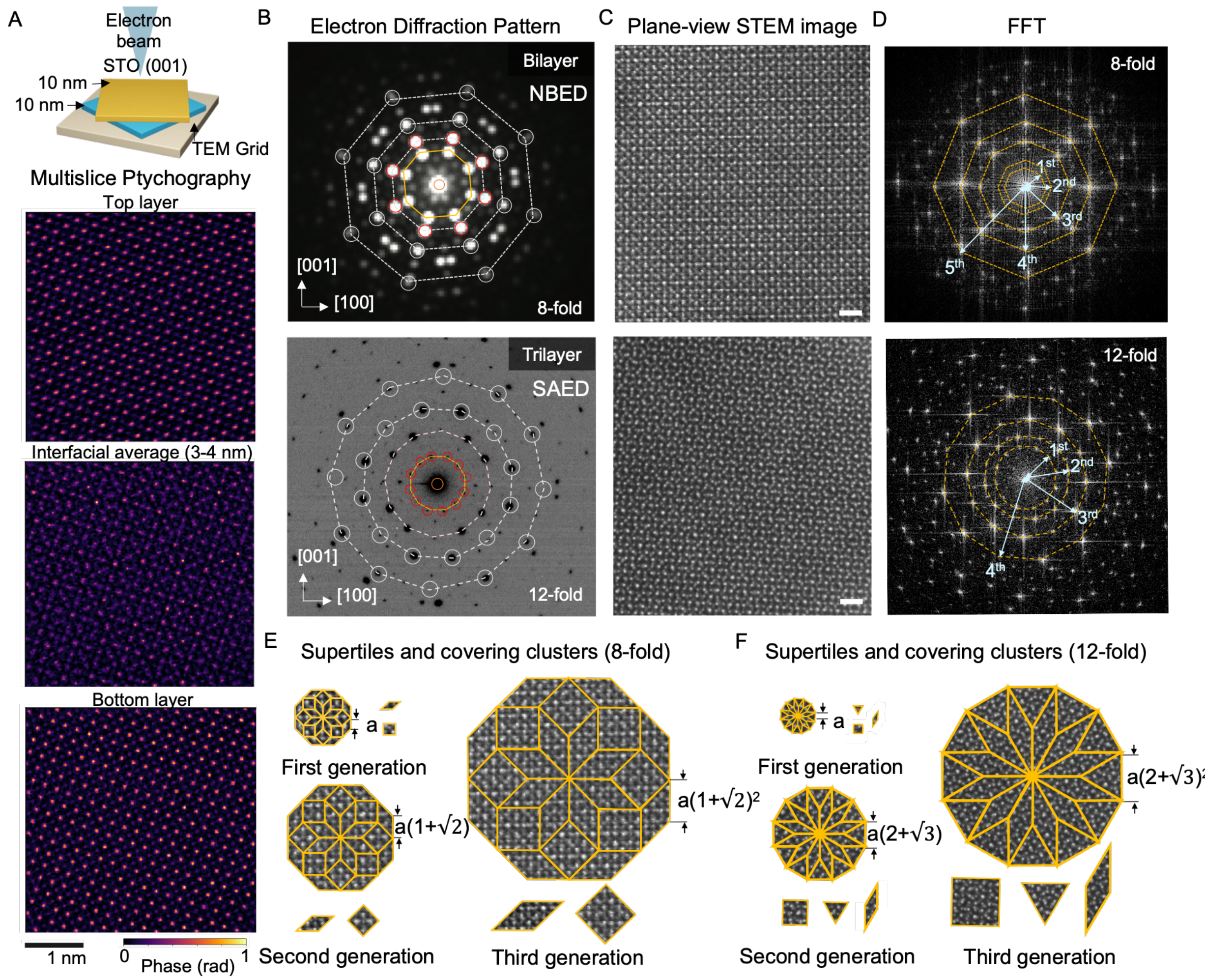

**Fig. 2.** Reciprocal-space and real-space signatures of 8-fold and 12-fold quasicrystalline order. (**A**) Sample schematic and multislice electron ptychography of a 45° twisted bilayer (10 nm/10 nm on a TEM grid), showing the top layer, interfacial average with 8-fold symmetry, and bottom layer. (**B**) Nano-beam electron diffraction (NBED) from a 45° twisted bilayer and selected-area electron diffraction (SAED) from a 30°/30° twisted trilayer, showing 8-fold and 12-fold symmetry, respectively. (**C**) Plane-view STEM images of the bilayer (10 nm/10 nm on a TEM grid) and trilayer (4.68, 5.46 and 3.12 nm on a TEM grid). Weaker intensity spots in trilayer arise from the thinner 3.12 nm bottom layer. Scale bar is 1 nm. (**D**) FFT patterns confirming 8-fold and 12-fold symmetry at atomic resolution. (**E,F**) Real-space quasicrystalline supertile hierarchies. First-, second-, and third-generation covering clusters scale as $a$, $a(1+\sqrt{2})$, and $a(1+\sqrt{2})^2$ for the 8-fold system, and as $a$, $a(2+\sqrt{3})$, and $a(2+\sqrt{3})^2$ for the 12-fold system.

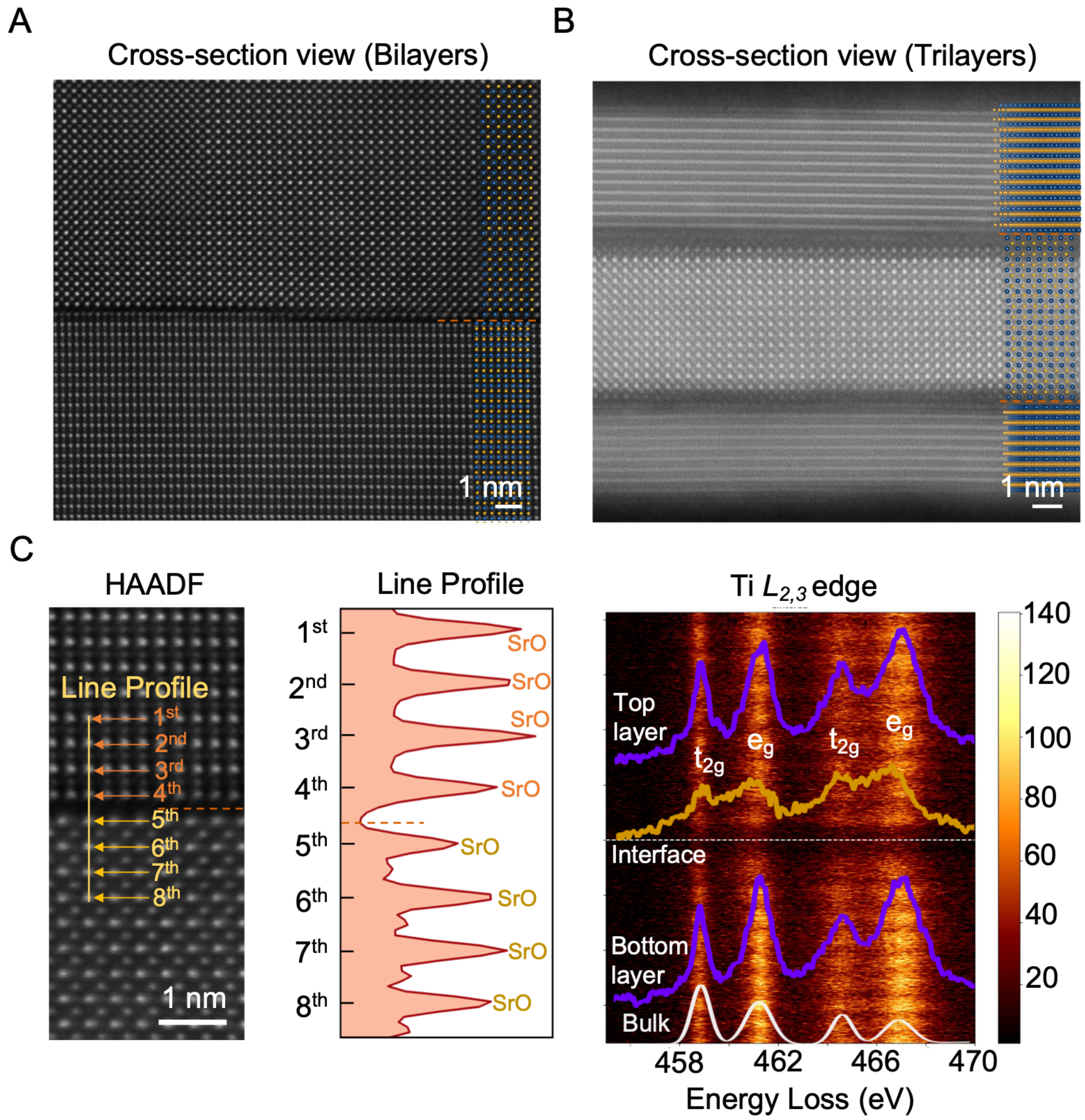

**Fig. 3.** Atomic-resolution interface imaging and electron energy loss spectroscopy. (**A**) High-resolution cross-sectional STEM image of a 45° twisted bilayer (30 nm STO/STO (001)) showing an atomically abrupt twist interface. (**B**) Cross-sectional STEM image of a 30°/30° twisted trilayer (4.68 nm/5.46 nm/3.12 nm on $SiO_2$/Si (001)) showing crystalline layer continuity and atomic step terraces. (**C**) HAADF-STEM image of a 45° twisted bilayer, corresponding intensity line profile across an SrO-SrO-terminated interface, and Ti $L_{2,3}$-edge EELS spectra acquired from bulk-like regions and the twist boundary. The spectra for bulk $Ti^{4+}$ in STO is taken from (*37*) is shown as a reference. The interface spectrum shows the emergence of a $Ti^{3+}$ contribution together with $Ti^{4+}$, consistent with local charge redistribution at the bonded twist boundary. Color scale indicates EELS intensity.

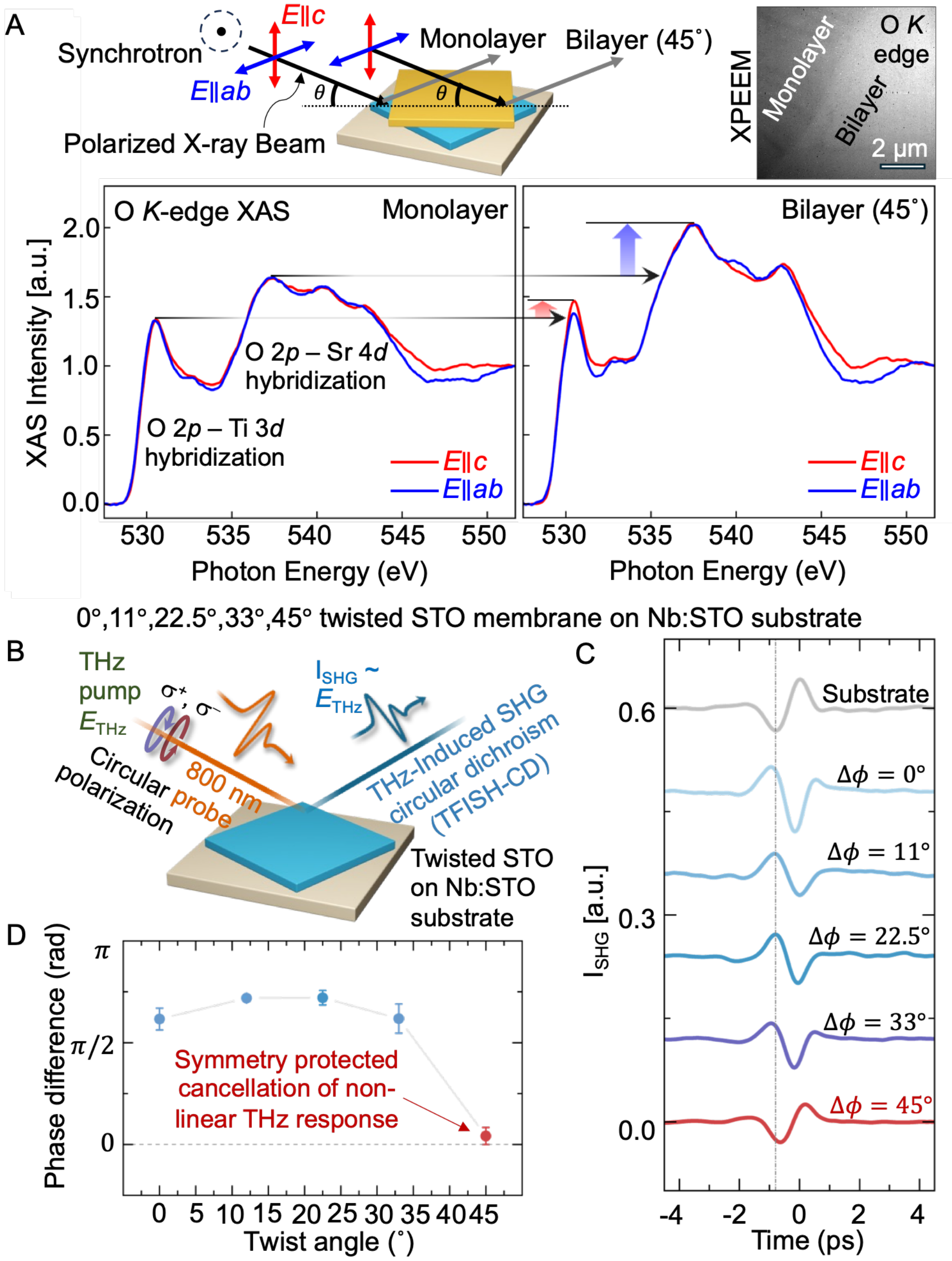

**Fig. 4.** Site-selective orbital reconstruction and nonlinear optical response at the quasicrystalline interface. (**A**) Top left: schematic of XAS/XLD measurements on a 45° twisted bilayer (1.5 nm STO/1.5 nm STO) on a solid substrate, showing polarization geometries predominantly sensitive to in-plane orbital character (E∥*ab*) and out-of-plane orbital character (E∥*c*). Top right: top-view XPEEM image at the O *K*-edge showing spatially resolved monolayer (ML; 1.5 nm) and bilayer (BL; Δ = 45°, 1.5 nm + 1.5 nm) regions within the same field of view. Bottom: O *K*-edge XAS spectra for the ML and quasicrystalline bilayer, with arrows indicating spectral-weight changes and dichroic contrast. (**B**) Schematic of TFISH-CD measurements at oblique incidence using circularly polarized 800 nm probe light on 0°, 11°, 22.5°, 33°, and 45° STO membranes on Nb:STO (001) (annealed). The THz field polarization is perpendicular to the optical plane. (**C**)

Twist-angle-dependent TFISH-CD signal. (**D**) TFISH-CD phase versus twist angle, showing the 45° interface as an optically cancelled twist boundary with symmetry-protected suppression of the nonlinear THz response.